\documentclass{article}
\usepackage{epsf}

\def\l{\langle}
\def\r{\rangle}

\begin{document}
\title{Application of exchange Monte Carlo \\
method to ordering dynamics}
\author{Yutaka Okabe\footnote{E-mail address: okabe@phys.metro-u.ac.jp}
\\
Department of Physics, Tokyo Metropolitan University, \\ 
Hachioji, Tokyo 192-0397, Japan}
\date{}

\maketitle
\begin{abstract}
We apply the exchange Monte Carlo method to 
the ordering dynamics of the three-state Potts model with 
the conserved order parameter.
Even for the deeply quenched case to low temperatures, we have 
observed a rapid domain growth;  we have proved the efficiency 
of the exchange Monte Carlo method for the ordering process.
The late-stage growth law has been found to be  $R(t) \sim t^{1/3}$ 
for the case of conserved order parameter of three-component system.
\end{abstract}


\vspace{2em}

The ordering dynamics in the spinodal decomposition has captured
a lot of attention \cite{Gunton83}.  
It is considered that the domain-size growth 
in the late stage is governed by an algebraic law, $R(t) \sim t^n$. 
The classical Lifshitz-Slyozov theory \cite{Lifshitz61}
gives the growth exponent $n$=1/3 
in the case of the spinodal decomposition of the conserved order 
parameter.  On the other hand, the late-stage ordering process of the 
nonconserved order parameter is described by the classical 
Lifshitz-Allen-Chan law, $n=1/2$ \cite{Lifshitz62,Allen79}.  
Although there has been controversy about the value of the growth 
exponent $n$ especially for the conserved order parameter case, 
the recent studies \cite{Jeppesen93,Laradji94} 
have revealed that those growth exponents are universal, 
that is, they do not depend on the space dimensionality 
or the number of the components of the order parameter.

In simulational studies of ordering process of phase separation, 
we often encounter the problem of slow dynamics.  
The problem of slow dynamics is found in the wide range of problems 
in computer simulations.  Other examples are 
the critical slowing down near the critical point and 
the slow dynamics due to randomness or frustration. 
There are several attempts to overcome slow dynamics in
the Monte Carlo (MC) simulation.  We may classify these attempts into
two categories.  The first one is the cluster algorithm, such as
that of Swendsen-Wang \cite{Swendsen87} and that of Wolff \cite{Wolff89}.  
The other one is the extended ensemble method. 
The multi-canonical method \cite{Berg91}, 
the simulated tempering \cite{Marinari92}, and the exchange MC 
method \cite{Hukushima96a} are examples of the second category.
The exchange MC method has been successfully applied to the
problem of spin glasses \cite{Hukushima96a,Hukushima96b}.

Usually, it is considered that the extended ensemble method 
can be used only for static problems because 
the extended ensemble affects the dynamics.  Is it always so?
In this paper, we apply the exchange MC method \cite{Hukushima96a}, 
which treats the exchange of replicas with different temperatures,
to the ordering problem, and test if we can discuss 
the ordering dynamics by using the exchange MC method.
We pay attention to the fact that the ordering phenomena 
after quenching are controlled by a zero-temperature fixed point, 
in the language of the renormalization group, irrespective of 
quenched temperature below the critical temperature, $T_c$. 
We also note that in the case of algebraic growth, the composite
of the growth law of different temperatures may become an algeraic 
one again in the leading order, which will be discussed later.

As a model system, we pick up the three-component system 
of the conserved order parameter.  The reason is as follows.  
The ordering dynamics of the three-component system 
(three-state Potts model) is slower compared to 
the two-component system (Ising model), and, moreover, 
the temporal growth of the conserved order parameter is slower 
than that of the nonconserved order parameter \cite{Jeppesen93,Laradji94}. 
Since it is a difficult job to determine the late-stage growth law 
of the conserved order parameter due to the slow dynamics, 
the three-component system of the conserved order parameter is 
suitable for testing the efficiency of the exchange MC method.

We perform the MC simulation of the three-state ferromagnetic Potts model 
\cite{Potts48,Wu82} on the square lattice, whose Hamiltonian is given by
\begin{equation}
 {\cal H}= -J \sum_{<i,j>} \delta (S_i, S_j), 
\end{equation}
where $S_j$ takes one of the three states, say, $a, b$ and $c$.
For the spin-update of the MC simulation, we employ the Kawasaki 
dynamics of nearest-neighbor pair exchange because we treat 
the conserved order parameter.

Here, we briefly review the exchange MC method \cite{Hukushima96a}. 
We treat a compound system which consists of $M$ replicas of the system.
The $m$-th replica is associated with the inverse temperature $\beta_m$.
We consider the extended ensemble, which is denoted by 
$\{X \} = \{X_1, X_2, \cdots X_M \}$.  
Then, the partition function 
of the compound system is given by
\begin{equation}
 Z={\rm Tr}_{\{X\}} \exp(-\sum_{m=1}^{M} \beta_m {\cal H} (X_m)) .
\end{equation}
To obtain an equilibrium distribution of the whole replicas, 
a replica exchange update process should be introduced. 
By considering the detailed balance condition, 
the transition probability of exchanging the $i$-th replica 
and the $j$-th replica may be chosen as 
\begin{equation}
 W(\{X_i,\beta_i;X_j,\beta_j\} \rightarrow \{X_i,\beta_j;X_j,\beta_i\}) 
  = 
 \left\{
 \begin{array}{ll}
 1  & \quad (\Delta \le 0) \\ e^{-\Delta} & \quad (\Delta > 0)
 \end{array}
 \right.
\end{equation}
where
\begin{equation}
 \Delta = -(\beta_i - \beta_j)({\cal H}_i-{\cal H}_j) .
\end{equation}
It should be noted that there is freedom of choice 
in inverse temperatures of replicas, $\beta_m$.
We should take account of conditions such that 
the replica exchange happens with a non-negligible probability 
for all adjacent pairs of replicas, and 
each replica moves around the whole temperature range 
in suitable Monte Carlo steps per spin (MCS).

We make a few comments on the coding of the MC simulation.  
We use the technique of the multispin coding
\cite{Bhanot86,Michael86,Kikuchi87,Kikuchi95}, 
where each bit within one word is assigned to the spin of 
different systems.  When using a 32-bit machine, one is
able to treat 32 systems simultaneously.  
For the case of the three-state Potts model, two words 
are needed to represent one set of spin states. 
This is different from the case of the Ising model, 
where spin states can be represented by one word.  
The multispin coding algorithm for the three-state Potts model 
obeying the Glauber dynamics has been already presented
\cite{Kikuchi95}.  Here we extend this algorithm to the case of 
the Kawasaki dynamics of nearest-neighbor pair exchange.  
As far as we know, the multispin coding algorithm 
for the Kawasaki dynamics has not been publicized so far 
even for the Ising model.  

\begin{figure}
\epsfxsize=6cm 
\centerline{\epsfbox{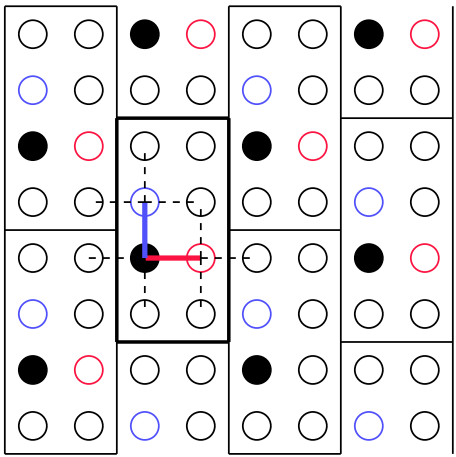}}
\vspace{2mm}
\caption{The schematic illustration of the sublattice 
decomposition for the MC of Kawasaki dynamics.}
\label{sublattice}
\end{figure}

It has been well known that the vectorization is possible 
if one decomposes the lattice into interpenetrating sublattices.  
This technique is effective for fast computation when using 
a vector computer.  In the case of the MC simulation of 
the Kawasaki dynamics, the sublattice decomposition into eight
sublattices is useful, which is illustrated in figure~\ref{sublattice}.
If a single spin represented by a black closed circle is picked up,
we can choose either a red circle or a blue circle as a pair spin. 
For the calculation of the local energy, we need the information 
on the states of spins connected by dotted lines.  All the spins 
concerned are independent of other black spins.  Thus, 
all the calculations become vectorized.  And this choice of 
sublattice decomposition is useful for the fast Fourier transform (FFT) 
calculation.

We have made simulations for the three-state ferromagnetic Potts model 
with the linear sizes, $L=$64, 128 and 256.  
We quench the system to desired temperatures, 
starting from the disordered state ($T=\infty$). 
Actually, we assign 16 systems of 32 multispin-coding systems
to the replica exchange MC calculation of 16 temperatures.  
The other 16 systems are used for the standard 
MC calculation.  For example, we consider 16 temperatures 
starting from 0.3 in units of $J$ with a separation of 0.04; 
that is, the highest temperature is 0.9.  The transition 
temperature of the three-state ferromagnetic Potts model 
on the square lattice is exactly known as 
$T_c = [\ln (1+\sqrt{3})]^{-1}$=0.99497 \cite{Wu82}. 
Since we deal with the ordering phenomena, we have chosen 
all the temperatures such as lower than $T_c$.  
The number of the spins which take each of three states
are fixed to be the same.  We make a trial for the replica exchange 
after one MCS for a single spin flip.  
Typical MCS are $8.0 \times 10^5$.  

\begin{figure}
\epsfxsize=10cm 
\centerline{\epsfbox{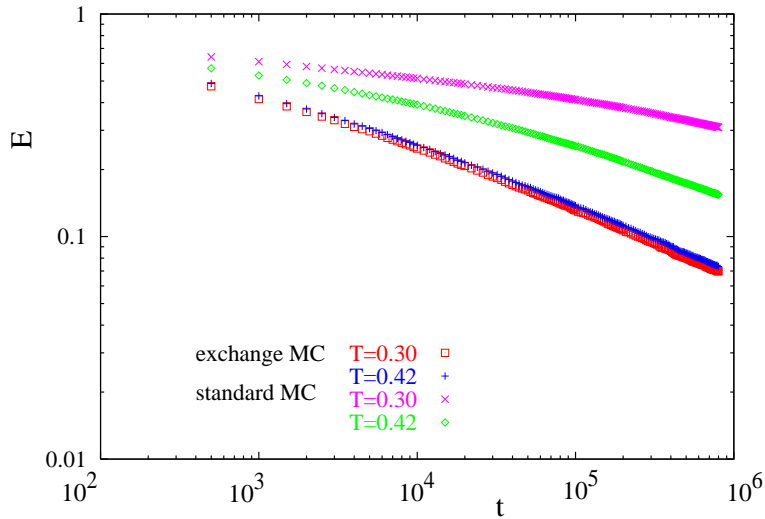}}
\caption{Time evolution of the excess energy, equation (\ref{excess}), 
for the three-state Potts model of the size $128 \times 128$ 
quenched to $T =$ 0.3 and 0.42.  The results for the exchange MC 
and the standard MC are compared.}
\label{energy}
\end{figure}

There are several ways to estimate the characteristic length scale 
of the domain size.  Binder and Stauffer \cite{Binder74} pointed out 
that the temporal change of the total energy from 
the equilibrium energy is appropriate for estimating the domain size;
that is, 
\begin{equation}
 R_E(t) = N [ \l {\cal H}(t) \r - \l {\cal H} \r_T]^{-1}
\end{equation}
where $\l {\cal H} \r_T$ is an equilibrium energy, and $\l \dots \r$ 
represents a sample average.
There are other quantities to estimate the domain size. 
For example, we may use the moment of the time-dependent 
structure factor $S(k)$.  We have calculated $S(k)$ by performing 
the FFT.  However, since the results are essentially the same, 
we only plot the time evolution of the excess energy 
\begin{equation}
 E = (\l {\cal H}(t) \r - \l {\cal H} \r_T)/N
\label{excess}
\end{equation}
in figure~\ref{energy}.  
The MCS is used for time $t$.  We compare the data of 
the exchange MC method and those of the standard MC method.
The system size is $128 \times 128$, and quenched temperatures are 
$T$ = 0.3 and 0.42.  The average has been taken over 16 samples 
for each data.

Comparing the purple and green curve in figure~\ref{energy}, 
we see that the domain growth becomes very slow at low temperatures 
for the standard MC.  It is because the thermal diffusion is 
not so frequent at low enough temperatures.  On the contrary, from 
the red and blue curves in figure~\ref{energy}, we find that 
the growth rate is faster for lower temperature in the case of 
the exchange MC because of the replica exchange process.

\begin{figure}
\epsfxsize=9cm 
\centerline{\epsfbox{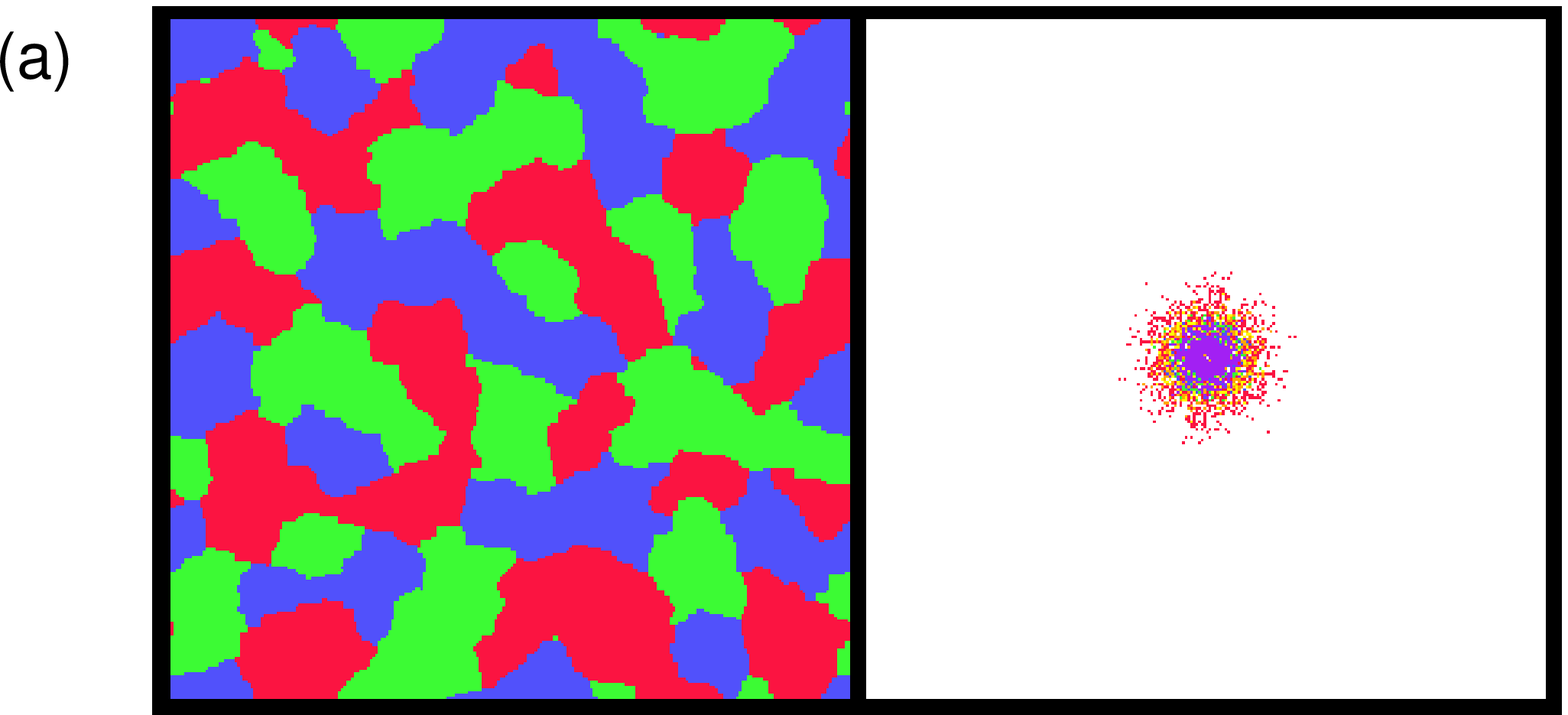}}
\vspace{10mm}
\epsfxsize=9cm 
\centerline{\epsfbox{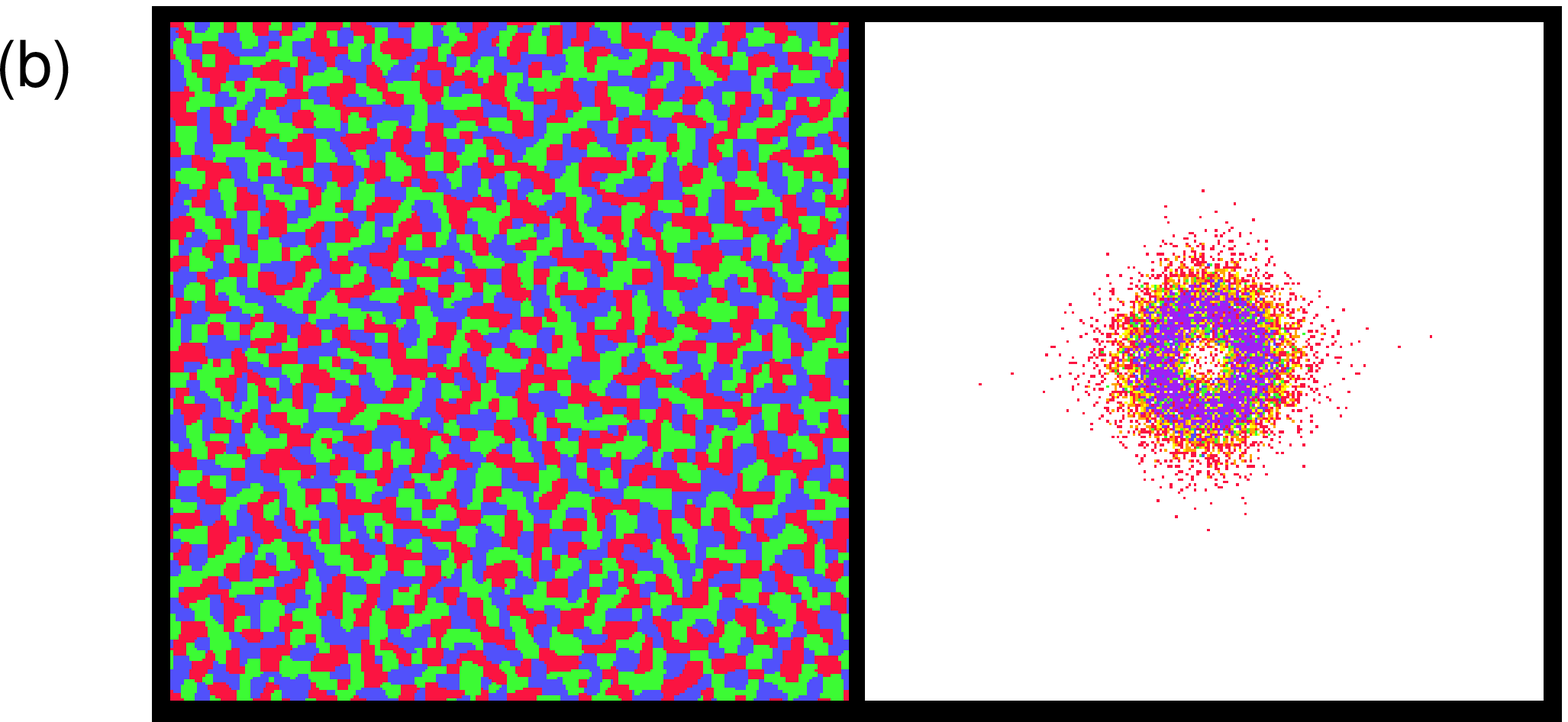}}
\caption{The real-space snapshot together with 
the structure factor $S(k)$ at $8.0 \times 10^5$ MCS 
(a) for exchange MC 
and (b) for standard MC.  The system size is $256 \times 256$  
and the quenching temperature is 0.3.}
\label{snapshot}
\end{figure}

Examples of the real-space snapshots and the corresponding 
structure factor $S(k)$ are given in figure~\ref{snapshot}, 
where the system size is $256 \times 256$ and 
the quenched temperature is $T=0.3$.  
The time is $8.0 \times 10^5$ MCS and the data for 
the exchange MC and for the standard MC are compared in (a) and (b), 
respectively.  The three colors, red, green and blue, are used 
for representing three states in the real-space snapshots.  
The vertical and horizontal axes for the structure factor is 
spanned from $-\pi$ to $\pi$ in units of the inverse lattice spacing.
We may see from figure~\ref{snapshot} that 
even for the deeply quenched case to low temperatures, 
we have observed a rapid domain growth
with the use of the exchange MC method.
In the case of $T=0.3$, to attain the same domain size, 
the calculation with the exchange MC method is about 200 times 
faster than that with the standard MC method at the cost of 
16 simultaneous calculations with different temperatures.  
We can use the data for all the temperatures, and the efficiency 
becomes more prominent for much lower temperatures.
%

\begin{figure}
\epsfxsize=10cm 
\centerline{\epsfbox{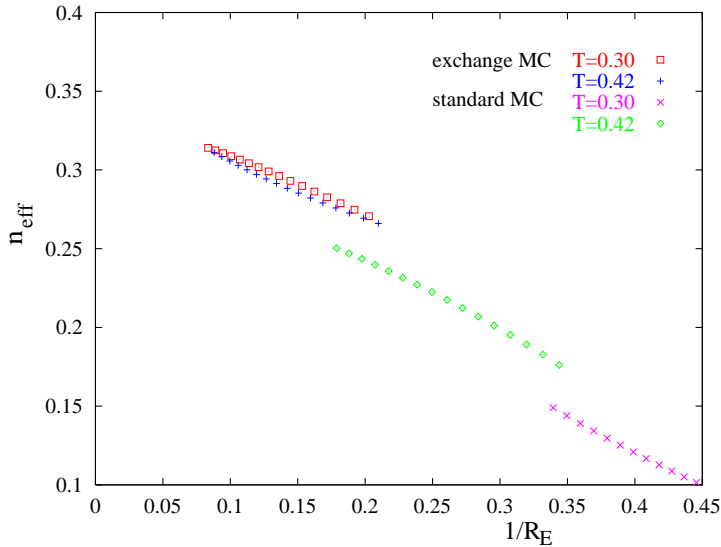}}
\caption{The effective exponent $n_{\rm eff}$ as a function of $1/R_E$.}
\label{effective}
\end{figure}

In order to estimate the growth exponent, 
it is convenient to consider the effective exponent 
defined by
\begin{equation}
 n_{\rm eff} = \frac{d \ln R (t)}{d \ln t}  .
\end{equation}
We can estimate the growth exponent $n$ by extrapolating 
$n_{\rm eff}$ in the limit of $t \rightarrow \infty$.
Following Huse \cite{Huse86}, we plot the effective exponent 
$n_{\rm eff}$ as a function of $1/R_E$ in figure~\ref{effective}.
We calculate $n_{\rm eff}$ from the data shown in figure~\ref{energy}.
We compare the data of the exchange MC method and those of the
standard MC.  From the data of the exchange MC method 
shown in figure~\ref{effective}, we may reliably estimate $n$ as 1/3, 
which is consistent with the conclusion of Ref.~\cite{Jeppesen93}. 
In the study of the standard MC simulation \cite{Jeppesen93}, 
the growth exponent has been estimated as 1/3 from long extrapolation 
process; the value of the effective exponent for the largest domain 
size was 0.23.  On the contrary, our value of $n_{\rm eff}$ 
by the use of the exchange MC simulation is 0.32.  Thus, the estimate 
of the exponent $n$ is more reliable.

Although the replica exchange dynamics is not considered to be related 
to a real one, we have found that a domain growth is controlled 
by a simple algebraic growth law, $R(t) \sim t^{1/3}$.  The value 
is consistent with a direct simulation \cite{Jeppesen93} 
for the same model.
In the standard MC the domain growth becomes very slow 
at low temperatures because of the lack of the thermal diffusion.
In order to study the late-stage dynamics in the standard MC, 
we often choose relatively high temperature, for example, $T=0.5T_c$. 
Then interfacial fluctuations become large. 
The advantage of using the exchange MC method is that 
we can have rapid growth of order together with small 
interfacial fluctuations.  The problem is whether the exchange 
process of replica modifies the dynamics or not.  The answer is 
yes generally.  However, for the problem of ordering phenomena, 
it works well even for the estimate of the growth exponent.  
Why can the exchange MC method simulate the dynamics of 
ordering phenomena?  The ordering phenomena 
are controlled by a zero-temperature fixed point, in the language 
of the renormalization group, irrespective of quenched temperature; 
thus the replicas associated with different (inverse) temperatures 
obey the same growth law.  Even if the growth process 
is described by the composite of the growth law of 
different temperatures, the resulting growth behavior 
again becomes an algebraic one,
\begin{equation}
 C_1 t^n + C_2 t^n + C_3 t^n + \cdots \propto t^n  .
\end{equation}

One comment should be made here on the choice of temperatures of 
replicas.  Although we have shown only the data for the 16 temperatures 
from 0.3 to 0.9 with the temperature separation of 0.04, other choices
of temperatures give essentially the same results.

In summary, we have tested the efficiency of the exchange MC method 
in the case of the ordering process by quenching. 
Even for the deeply quenched case at low temperatures, 
we have observed a rapid domain growth. 
Although the dynamics including the exchange process is not a simple one, 
a domain growth has been found to be controlled by a simple 
algebraic growth law, $R(t) \sim t^{1/3}$, for the case of 
conserved order parameter of three-component system.
It is interesting to apply this method to 
more complicated problems.  The effect of surfactants 
in the phase separation dynamics is now studied by using 
this method \cite{Ito99}.

The author would like to thank T. Kawakatsu, T. Ito, K. Hukushima 
and K. Nemoto for valuable discussions. 
This work was supported by a Grant-in-Aid for Scientific Research 
from the Ministry of Education, Science, Sports and Culture, Japan.
The computation in this work has been done using the facilities of 
the Supercomputer Center, Institute for Solid State Physics, 
University of Tokyo.

\end{document}